\documentclass[11pt,reqno]{article}
\usepackage{amsfonts}
\usepackage{amsmath}
\usepackage{amsbsy}
\usepackage{graphicx}
\usepackage{amssymb,latexsym}
\usepackage{graphicx}
\usepackage{float}
\usepackage{caption}
\usepackage{subcaption}
\numberwithin{equation}{section}

\begin{document}

\begin{titlepage}

\title{Bianchi-Type Black Holes in Higher-Dimensional $R^2$ Gravity}

\author{ Se\c{c}il \c{S}entorun \footnote{E.mail:secilo@eskisehir.edu.tr} \\
{\small Department of Physics, Eskisehir Technical University, 26470 Eski\c{s}ehir, Turkey}}

\date{ }

\maketitle

\bigskip

\begin{abstract}

\noindent In this study, we investigate higher-dimensional static and stationary Bianchi-type solutions in an $R^2$ gravity theory using the formalism of exterior differential forms. We construct a broad class of exact higher-dimensional Bianchi-type solutions in an $R^{2}$-corrected gravity theory. It is shown that the theory admits hyperscaling violating black hole solutions associated with homogeneous but anisotropic horizon geometries, encompassing both static and stationary configurations. We further analyze some thermodynamic properties of these backgrounds and demonstrate that higher-curvature corrections give rise to unconventional features. In particular, we show that within the critical sector, the solutions possess a finite Hawking temperature despite an identically vanishing Wald entropy, signaling an unusual thermodynamic regime. These configurations arise at the critical point of the theory, where the effective gravitational coupling vanishes and the space of admissible constant-curvature geometries is enlarged.

\end{abstract}

\end{titlepage}

\section{Introduction}

Exact black hole solutions in higher-dimensional gravity theories play a central role in understanding the interplay between geometry, dynamics, and thermodynamics beyond Einstein gravity. In particular, gravitational theories supplemented by higher-curvature corrections arise naturally as effective descriptions of quantum gravity and string-inspired models. Among these, theories containing a quadratic curvature term such as $R^2$ gravity provide a minimal extension of general relativity in which nontrivial modifications to black hole structure and thermodynamics can be studied analytically \cite{boulware}.

A defining feature of higher-curvature gravity theories is their impact on black hole thermodynamics. The presence of quadratic curvature terms typically leads to deviations from the standard area law for entropy and introduces explicit dependence of thermodynamic quantities on coupling constants and spacetime dimensionality. These effects have been extensively discussed in the literature, emphasizing the importance of constructing exact solutions in order to properly assess their physical implications \cite{myers}.

While most known black hole solutions possess horizons of maximal symmetry, it has become increasingly clear that more general homogeneous but anisotropic horizon geometries may also arise in higher-dimensional settings. Such geometries are naturally classified by the Bianchi scheme, which provides a systematic framework for describing homogeneous spaces with distinct symmetry properties \cite{ryan}. Black holes with Bianchi-type horizon geometries therefore constitute a natural generalization of the standard isotropic solutions and allow for a richer geometric structure.

Higher-dimensional black holes endowed with Bianchi-type horizon geometries have been constructed in various gravitational theories, demonstrating that both static and stationary configurations can consistently support homogeneous anisotropy on the horizon \cite{hassaine}. In the presence of higher-curvature corrections, additional constraints on the admissible horizon geometries may arise, leading to qualitative modifications of the solution space. In particular, quadratic curvature terms have been shown to impose nontrivial restrictions on horizon topology and geometry, thereby affecting both the existence and properties of black hole solutions \cite{dotti}.

From a thermodynamic perspective, anisotropic horizon geometries combined with higher-curvature corrections may lead to unconventional behavior. The dependence of horizon temperature and entropy on spacetime dimension, rotation parameters, and curvature couplings can differ substantially from that observed in Einstein gravity. Such deviations provide valuable insights into the role of geometric anisotropy and higher-order curvature effects in black hole thermodynamics.

Motivated by these considerations, the present work investigates higher-dimensional static and stationary Bianchi-type solutions in an $R^2$-corrected gravity theory formulated using the language of exterior differential forms. This formalism provides a natural and efficient description of the geometric and variational structure of the theory, particularly in the presence of torsionless, metric-compatible connections. Both static and stationary configurations with hyperscaling violating configurations associated with homogeneous anisotropic horizon geometries are analyzed, with particular emphasis on their geometric properties and thermodynamic behavior. Selected thermodynamic quantities are evaluated, and their dependence on the higher-curvature coupling and spacetime dimensionality is discussed. The results presented in this work indicate that quadratic curvature corrections can support a broad class of anisotropic black hole geometries with Bianchi-type horizon structures in arbitrary spacetime dimensions. In this sense, the solutions obtained here may be viewed as natural extension of previously studied anisotropic black hole configurations to a wider geometric setting that includes both static and stationary hyperscaling–violating configurations. The results illustrate how anisotropy, dimensionality, and higher-curvature effects combine to yield nontrivial black hole thermodynamics in these backgrounds.

An interesting feature of the solutions presented here is that the horizon geometry belongs to the Bianchi class while the spacetime simultaneously exhibits hyperscaling–violating anisotropic scaling behavior. To the best of our knowledge, exact black hole configurations combining these ingredients in higher–dimensional $R^2$ gravity have not been systematically explored. The present construction therefore enlarges the family of known anisotropic solutions in higher–curvature gravity and provides a concrete framework in which homogeneous but anisotropic horizon geometries coexist with hyperscaling violation.

The paper is organized as follows. In Section 2, we introduce the mathematical background used in this study. Section 3 describes the action and variational procedure used to derive the field equations of the $R^2$ -corrected gravity model. In Section 4, we construct static and stationary Bianchi-type solutions. Section 5 is devoted to the thermodynamic analysis of these configurations. Finally, Section 6 summarizes our findings and outlines possible directions for future work.

\section{Mathematical Background}

Throughout this work, we employ the language of exterior differential forms on an $n$-dimensional pseudo-Riemannian manifold $M$ equipped with a metric tensor $g$. The geometry is described by an orthonormal coframe $\{e^{a}\}$ satisfying
\begin{equation}
g=\eta_{ab}\,e^{a}\otimes e^{b},
\end{equation}
where $\eta_{ab}$ denotes the Minkowski metric. The corresponding Levi--Civita connection $1$-forms $\omega^{a}{}_{b}$ are determined by Cartan's first structure equation
\begin{equation}
de^{a}+\omega^{a}{}_{b}\wedge e^{b}=0 ,
\end{equation}
together with the metric compatibility condition
\begin{equation}
\omega_{ab}+\omega_{ba}=0 .
\end{equation}
The curvature $2$-forms are defined through Cartan's second structure equation
\begin{equation}
R^{a}{}_{b}=d\omega^{a}{}_{b}+\omega^{a}{}_{c}\wedge\omega^{c}{}_{b}.
\end{equation}
From the curvature $2$-forms one obtains the Ricci $1$-forms
\begin{equation}
P_{a}=\iota_{b}R^{b}{}_{a},
\end{equation}
and the scalar curvature
\begin{equation}
R=\iota^{a}P_{a}.
\end{equation}
Here $\iota_{a}$ denotes the interior product with respect to the frame vector dual to $e^{a}$. The Hodge dual operator associated with the metric is denoted by $\ast$ and acts on differential forms in the standard way. Throughout the paper we adopt the conventions and notation commonly used in the differential-form formulation of gravitational theories. 

\section{Field Equations from Action}

The gravitational model considered in this work is based on a quadratic modification of Einstein gravity in which the action depends only on the Ricci scalar and a cosmological constant term. In the language of exterior
differential forms, the action is generated by the Lagrangian $n$-form
\begin{equation}
L=\left(\frac{1}{2} \left( \alpha R^2 + R\right) + \Lambda \right) \ast 1,
\label{eq:Lagrangian}
\end{equation}
where $R$ denotes the scalar curvature, $\Lambda$ is the cosmological constant, $\alpha$ is the quadratic curvature coupling parameter and $\ast 1$ represents the invariant volume element.
The variational formulation of this model can be developed by treating the connection and co-frame 1-forms as the fundamental dynamical variables while assuming a torsion-free and metric-compatible Levi-Civita connection. The detailed derivation of the field equations within the exterior-form formalism for the same $R^{2}$ gravity model was presented in our previous study devoted to higher-dimensional anisotropic black hole solutions \cite{sentorun}. For completeness, we only quote the resulting equations required for the present analysis.
Variation of the action yields
\begin{equation}
\left(\alpha R+\frac{1}{2}\right)R^{ab}\wedge \ast(e_a\wedge e_b\wedge e_c)+\left(\Lambda-\frac{\alpha}{2}R^2\right)\ast e_c+2\alpha D(\iota_c \ast dR)=0 ,
\label{eq:fieldeq}
\end{equation}
where $D$ denotes the exterior covariant derivative associated with the Levi-Civita connection and $\iota_c$ is the interior product operator.

The Bianchi-type configurations investigated in this work belong to the constant-curvature sector of the theory. Consequently,
\begin{equation}
dR=0 ,
\label{eq:constantR}
\end{equation}
and the derivative contribution in Eq.~(\ref{eq:fieldeq}) vanishes identically. The field equations therefore simplify to
\begin{equation}
\left(\alpha R+\frac{1}{2}\right)R^{ab}\wedge\ast(e_a\wedge e_b\wedge e_c)+\left(\Lambda-\frac{\alpha}{2}R^2\right)\ast e_c=0 .
\label{eq:reducedeq}
\end{equation}
Contracting Eq.~(\ref{eq:reducedeq}) gives the scalar relation
\begin{equation}
(1+2\alpha R)R-4\Lambda=0 .
\label{eq:trace}
\end{equation}
Among the admissible branches of the theory, a distinguished sector is obtained by imposing
\begin{equation}
1+2\alpha R=0 .
\label{eq:critical}
\end{equation}
Substitution of Eq.~(\ref{eq:critical}) into Eq.~(\ref{eq:trace}) immediately yields
\begin{equation}
R=-4\Lambda ,
\label{eq:Rcritical}
\end{equation}
together with
\begin{equation}
\alpha=\frac{1}{8\Lambda}.
\label{eq:alphacritical}
\end{equation}
This parameter choice corresponds to a critical point of the theory at which the effective gravitational coupling
\begin{equation}
f'(R)=1+2\alpha R
\end{equation}
vanishes. Similar degeneracies have been discussed previously in several higher-derivative gravity models and are known to enlarge the space of admissible constant-curvature solutions \cite{lu1, kehagias}.

The static and stationary Bianchi-type geometries constructed in the following sections are obtained within this critical sector. In this regime, the Einstein metrics satisfying Eq.~(\ref{eq:Rcritical}) automatically solve
the quadratic field equations, thereby allowing a wider class of anisotropic backgrounds than in the generic parameter domain of the theory.

The present work builds upon our previous analysis of higher-dimensional anisotropic backgrounds \cite{sentorun} by incorporating the full Bianchi-type symmetry classification.

\section{Bianchi type solutions}

\noindent 

Motivated by the analysis presented in \cite{lu2}, where five-dimensional $R^2$-corrected gravity was studied in the context of homogeneous anisotropic geometries and nine classes of Bianchi-type solutions, we investigate whether
analogous Bianchi-type configurations persist in the higher-dimensional $R^2$ gravity model introduced in Section 3 and whether they admit black-hole generalizations with hyperscaling-violating asymptotics.

\subsection{Bianchi type static solutions}

\noindent To construct anisotropic black hole configurations compatible with the homogeneous horizon geometries classified by the Bianchi scheme, we begin with the general metric ansatz
\begin{equation}
ds^2=-e^{2\beta_t r}dt^2+dr^2+\eta_{ij} e^{(\beta_i+\beta_j)r}dx^i dx^j , 
\end{equation}
where $\beta_t$ and $\beta_i$ are constant parameters and $\eta_{ij}$ is a constant, coordinate-independent matrix. This class of metrics admits a generalized scaling symmetry,
\begin{equation}
r \rightarrow r+\epsilon  ,\quad
t \rightarrow t e^{-\beta_t \epsilon}  ,\quad
\omega^i \rightarrow \omega^i e^{-\beta_i \epsilon}  ,
\end{equation}
with $\omega^i = dx^i$ for $i=1,2,3,\ldots,(n-2)$. Within the critical sector defined by Eqs. (\ref{eq:Rcritical}) and (\ref{eq:alphacritical}), the reduced field equations (\ref{eq:reducedeq}) admit a static vacuum Bianchi-type solution of the form
\begin{equation} \label{331_1}
ds^{2}= -e^{2\beta_{0} r} dt^{2}+dr^{2} + e^{2\beta_{2} r}dx_{i} dx^{i},
\end{equation}
provided that the cosmological constant satisfies
\begin{equation} \label{331_2}
\Lambda=\frac{1}{4}\left(2\beta_0^2+2(n-2)\beta_0\beta_2+(n-2)(n-1)\beta_2^2\right),
\end{equation}
together with Eqs. (\ref{eq:Rcritical}) and (\ref{eq:alphacritical}). The vacuum solution (\ref{331_1}) can be considered as the initial geometry. Allowing for a nontrivial radial metric function leads to the more general black-hole configuration
\begin{equation} \label{331_4}
ds^{2}=-e^{2\beta_{0} r} f^{2}(r) dt^{2}
+e^{2\beta_{1} r}\frac{dr^{2}}{f^{2}(r)}
+e^{2\beta_{2} r}\frac{dx_{i} dx^{i}}{\left(1+\kappa\frac{\rho^2}{4}\right)^2},
\end{equation}
where $\beta_0$, $\beta_1$, and $\beta_2$ are constants. Substituting the ansatz (\ref{331_4}) into the reduced field equations (\ref{eq:reducedeq}) reduces the problem to a second-order ordinary differential equation for
$f^2(r)$, whose general solution is
\begin{equation} \label{331_5}
f^2(r)=c_{1} e^{p_{+}r}+c_{2} e^{p_{-}r}+c_{3}e^{2\beta_{1}r}+c_4 e^{2(\beta_1-\beta_2)r} .
\end{equation}
The exponents $p_{\pm}$ take the form
\begin{equation} \label{331_6}
p_{\pm}=\frac{-\left(3\beta_{0}-\beta_{1}+2(n-2)\beta_{2}\right)
\pm \sqrt{\left(\beta_{0}+\beta_{1}\right)^{2}
+4(n-2)\beta_{2}(\beta_{0}+\beta_{1}-\beta_2)}}{2},
\end{equation}
while the coefficients $c_3$ and $c_4$ are determined as
\begin{equation} \label{331_7}
c_{3}=\frac{4\Lambda}{4 \beta_1^2+2 d_1 \beta_1+d_2},
\end{equation}
\begin{equation} \label{331_8}
c_{4}=\frac{(n-2)(n-3)\kappa}{4(\beta_1-\beta_2)^2+2 d_1(\beta_1-\beta_2)+d_2},
\end{equation}
with
\begin{equation} \label{331_9}
d_1=3\beta_0-\beta_1+2(n-2)\beta_2,
\end{equation}
\begin{equation} \label{331_10}
d_2=2\beta_0(\beta_0-\beta_1)+2(n-2)(\beta_0-\beta_1)\beta_2+(n-2)(n-1)\beta_2^2.
\end{equation}
The coefficients $c_3$ and $c_4$ are completely determined by the cosmological constant and the curvature of the transverse space, whereas $c_1$ and $c_2$ remain arbitrary integration constants associated with the homogeneous sector of the radial field equation. The curvature index $\kappa=-1,0,+1$ specifies the geometry of the $(n-2)$-dimensional transverse space, corresponding to hyperbolic, flat, and spherical geometries, respectively. The quantity $\rho$ denotes the radial coordinate on this subspace, defined by $\rho^2=x_i x^i$, with the index $i$ taking values $2,3,\ldots,(n-2)$.

\noindent For the metric function to be real, the parameter constraint (\ref{331_6}) requires
\begin{equation} \label{331_11}
\left(\beta_{0}+\beta_{1}\right)^{2}
+4(n-2)\beta_{2}(\beta_{0}+\beta_{1}-\beta_2)\geq 0.
\end{equation}

\noindent Under this condition, the solution (\ref{331_4}) can be interpreted as a Bianchi-type static black hole spacetime. A particularly simple subclass arises when the characteristic exponents
coincide. In this case,
\begin{equation} 
\beta_2=\frac{1}{2} \left( \beta_0 + \beta_1 \pm \frac{\sqrt{(n-1)(n-2)(\beta_0+\beta_1)^2}}{(n-2)}\right)
\end{equation}
ensures that the exponents $p_+$ and $p_-$ are equal. 

\noindent The horizon is located at $r=r_{+}$, defined as the largest root of $f^{2}(r_{+})=0$. A broad class of solutions is obtained for parameter choices satisfying Eqs. (\ref{331_6})–(\ref{331_10}), demonstrating the richness of the solution space within the $R^{2}$-corrected theory. From a physical perspective, the parameters $\beta_0$, $\beta_1$, and $\beta_2$ control the anisotropic scaling between the temporal, radial, and transverse directions, thereby determining the causal structure and horizon geometry of the spacetime. The curvature index $\kappa$ further distinguishes the horizon topology, allowing for hyperbolic, planar, or spherical black hole configurations, which in turn influence the thermodynamic properties of the solutions.

\noindent To illustrate that the algebraic constraints derived above admit physically acceptable solutions, we consider representative parameter sets satisfying Eqs. (\ref{331_6})–(\ref{331_11}). Figure 1 displays the behavior of the metric function for a parameter choice admitting a positive real root, corresponding to a black hole with an event horizon. For comparison, Figure 2 shows a representative parameter set for which the metric function remains strictly positive throughout the physical region $r>0$, indicating the absence of an event horizon and therefore a horizonless configuration. These examples confirm that the solution family contains both black hole and horizonless configurations depending on the choice of integration constants.

\begin{figure}[H]
\centering
\includegraphics[width=0.7\textwidth]{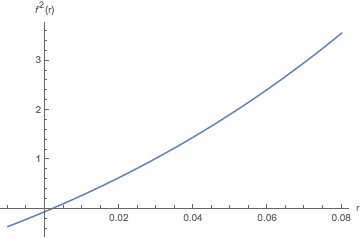}
\caption{Metric function of a static and spherical Bianchi type black hole solution admitting an event horizon located at $r_+=0.00177418$. Here we take $c_1=6.304895$, $c_2=-6.772287$, $n=5$, $\kappa=1$, $\Lambda=3.122499$, $\beta_0=-0.303521$, $\beta_1=-2.391589$ and $\beta_2=-1.626807$.}
\label{static_horizon}
\end{figure}

\begin{figure}[H]
\centering
\includegraphics[width=0.7\textwidth]{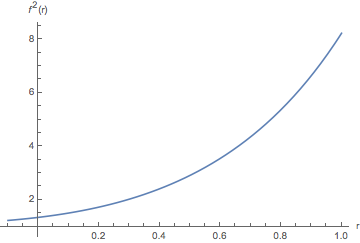}
\caption{Metric function of a static and spherical Bianchi type black hole solution featuring no positive real roots, establishing the presence of a horizonless configuration. Here, we take $c_1=0.90131$, $c_2=0.338498$, $n=5$, $\kappa=1$, $\Lambda=-1.614476$, $\beta_0=-0.422508$, $\beta_1=-2.023078$ and $\beta_2=-0.262018$.}
\label{static_naked}
\end{figure}

The metric (\ref{331_4}) can be expressed in an exponential radial gauge, which obscures its scaling properties at first sight. However, by performing a simple redefinition of the radial coordinate, $u=e^r$, one finds that the geometry can be rewritten in a hyperscaling–violating form:
\begin{equation}
ds^2=-u^{-2\theta/n} \left( -u^{2z} f(u) dt^2+\frac{du^2}{u^2 f(u)}+\frac{u^2 dx_i dx^i}{\left( 1+\kappa \frac{\rho^2}{4}\right)^2}\right) \, .
\end{equation}

\noindent In this parametrization, the effective dynamical and hyperscaling violation exponents are identified as  
\begin{equation}
z=\beta_0 +\beta_2 \quad , \quad \theta = n(1-\beta_2) \quad \mbox{and} \quad \beta_1=\beta_2 -1\, ,
\end{equation}
where $z$ and $\theta$ denote the dynamical and hyperscaling-violation exponents, respectively.

\subsection{Bianchi type stationary solutions}

Unlike the static case, the vacuum field equations do not admit a purely stationary Bianchi-type background compatible with the critical constant-curvature sector. Nevertheless, rotating geometries can still be
constructed by introducing an off-diagonal metric component that mixes the temporal and angular directions:
\begin{equation} \label{332_1}
ds^{2}=-e^{2\beta_{0}r} f^{2}(r) dt^{2}+e^{2\beta_{1}r}\left(d\phi+ \frac{\omega}{r^{2}} d t \right)^{2}+e^{2 \beta_{2}r}\frac{dr^{2}}{f^{2}(r)}+e^{2\beta_{3}r} \frac{dx_{i} dx^{i}}{\left(1+\kappa \frac{\rho^{2}}{4}\right)^{2}} . 
\end{equation}
Here, $f(r)$ is a radial metric function, $\beta_{0}$, $\beta_{1}$, $\beta_{2}$, and $\beta_{3}$ are constant parameters characterizing the anisotropic scaling of the spacetime, and the off-diagonal term proportional to $\omega$ generates frame dragging and represents the rotational deformation of the underlying static geometry. The curvature index $\kappa=-1,0,+1$ specifies the geometry of the $(n-3)$-dimensional transverse space, while $\rho$ represents its radial coordinate, defined by $\rho^{2}=x_{i}x^{i}$. In this configuration, the index $i$ runs over $1,2,\ldots,(n-3)$.

\noindent Substituting the stationary ansatz (\ref{332_1}) into the reduced field equations (\ref{eq:reducedeq}) yields a radial differential equation for $f^2(r)$. Solving the associated characteristic equation gives
\begin{equation} \label{332_2}
f^2(r)=c_{1} e^{p_{+}r}+c_{2} e^{p_{-}r}+c_{3}e^{2\beta_{2}r}+c_4 e^{2(\beta_2 -\beta_3)r}+c_5\frac{e^{2(\beta_1 -\beta_0)r}}{ r^4} , 
\end{equation}
where the characteristic exponents $p_{\pm}$ are given by
\begin{eqnarray} \label{332_3}
p_{\pm}&=& \frac{1}{2}\Big(-\left(2(n-3)\beta_{3}+3\beta_{0}+2\beta_{1}-\beta_{2}\right) \nonumber \\
&&\pm \left({\beta_{0}}^{2}-4{\beta_{1}}^{2}+{\beta_{2}}^{2} + 2\beta_{0}(2\beta_{1}+\beta_{2})+4\beta_{1}\beta_{2}+4(n-3)\beta_{3}(\beta_{0}+\beta_{2}-\beta_{3})\right)^{1/2}\Big) .\nonumber \\
\end{eqnarray}
The remaining coefficients entering Eq.~(\ref{332_2}) are determined as
\begin{equation} \label{332_4}
c_{3}=\frac{4\Lambda}{4 \beta_2^2+2 d_1 \beta_2+d_2} , 
\end{equation}
\begin{equation} \label{332_5}
c_{4}=\frac{(n-3)(n-4)\kappa}{4( \beta_2-\beta_3)^2+2 d_1( \beta_2-\beta_3)+d_2} , 
\end{equation}
\begin{equation} \label{332_6}
c_{5}=\frac{\omega^2}{10} , 
\end{equation}
where the auxiliary parameters are defined by
\begin{equation} \label{332_7}
d_1=3\beta_0+2\beta_1-\beta_2+2(n-3)\beta_3 , 
\end{equation}
\begin{equation} \label{332_8}
d_2=2\left(\beta_0^2+\beta_0\beta_1+\beta_1^2-(\beta_0+\beta_1)\beta_2\right)+2(n-3)(\beta_0+\beta_1-\beta_2)\beta_3+(n-3)(n-2)\beta_3^2  .
\end{equation}
\noindent Here, $c_1$, $c_2$, $c_3$, $c_4$, and $c_5$ are integration constants arising from the radial field equations. The coefficient $c_5$ originates entirely from the rotational sector and vanishes in the static limit $\omega \rightarrow 0$. The curvature index $\kappa=-1,0,+1$ characterizes the geometry of the $(n-3)$-dimensional transverse subspace, while $\rho$ denotes its radial coordinate, defined by $\rho^2=x_i x^i$.

\noindent Stationary Bianchi-type solutions in $n$-dimensional spacetime are obtained provided that Eqs. (\ref{eq:Rcritical}) and (\ref{eq:alphacritical}) holds together with
\begin{equation} \label{332_10}
4( \beta_1-\beta_0)+d_1=0 , 
\end{equation}
\begin{equation} \label{332_11}
4( \beta_1-\beta_0)^2+2 d_1( \beta_1-\beta_0)+d_2 =0 .
\end{equation}
Furthermore, the requirement that the metric function be real imposes the following condition
\begin{equation} \label{332_12}
{\beta_{0}}^{2}-4{\beta_{1}}^{2}+{\beta_{2}}^{2} + 2\beta_{0}(2\beta_{1}+\beta_{2})+4\beta_{1}\beta_{2}+4(n-3)\beta_{3}(\beta_{0}+\beta_{2}-\beta_{3}) \geq 0 .
\end{equation}
\noindent This solution demonstrates that higher-curvature terms allow for nontrivial rotating Bianchi-type geometries even in the absence of stationary vacuum configurations. Moreover, the exponents $p_+$ and $p_-$ are equal by
\begin{equation}
\beta_3=\frac{1}{2}\left( (\beta_0 + \beta_2)\pm \frac{\sqrt{(n-3) \left( (n-2)(\beta_0+\beta_1)^2+4 \beta_1 (\beta_0 - \beta_1 +\beta_2) \right)}}{(n-3)}\right) \, .\\
\end{equation}

\noindent Moreover, setting $\omega=0$ removes the rotational sector of the metric and yields the corresponding static Bianchi-type solution, confirming that the stationary spacetime represents a smooth rotational deformation of the static geometry.

\noindent As in the static case, the solution (\ref{332_2}) can be interpreted as a Bianchi-type stationary black hole spacetime, whose horizon geometry is determined by the value of the curvature index $\kappa$. The parameter space admits a wide class of solutions, provided that the constants appearing in the metric function satisfy the corresponding algebraic constraints. The largest root of the equation $f^2(r)=0$ determines the location of the event horizon $r=r_+$. The curvature invariants remain finite at this surface, while the true curvature singularity occurs only at smaller radial values. Moreover, by setting the rotation parameter $\omega=0$ and appropriately redefining the remaining constants, the stationary configuration smoothly reduces to the Bianchi-type static solutions, as expected.

\begin{figure}[H]
\centering
\includegraphics[width=0.7\textwidth]{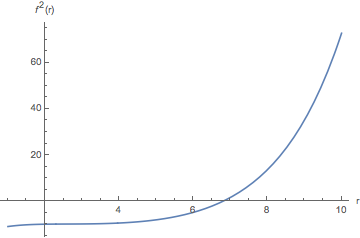}
\caption{Metric function of a stationary and spherical Bianchi type black hole solution admitting an event horizon located at $r_+=6.79995$. Here we take $c_1=-7.205257$, $c_2=-5.365793$, $n=5$, $\kappa=1$, $\Lambda=-4.858181$, $\beta_0=0.99813$, $\beta_1=0.31091$, $\beta_2=0.056889$, $\beta_3=-0.20261$ and $\omega=2.292586$.}
\label{stationary_horizon}
\end{figure}

\begin{figure}[H]
\centering
\includegraphics[width=0.7\textwidth]{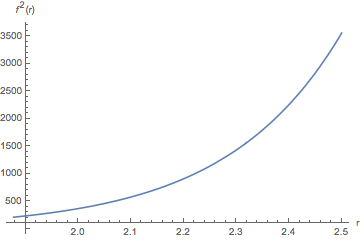}
\caption{Metric function of a stationary and spherical Bianchi type black hole solution featuring no positive real roots, demonstrating the presence of a horizonless configuration. Here, we take $c_1=8.242297$, $c_2=9.907065$, $n=5$, $\kappa=1$, $\Lambda=2.795582$, $\beta_0=3.772051$, $\beta_1=-2.333826$, $\beta_2=2.280321$, $\beta_3=5.013833$ and $\omega=1.240693$.}
\label{stationary_naked}
\end{figure}

\noindent As in the static case, the stationary solution also admits physically distinct configurations depending on the integration constants. Representative examples are presented in Figs. \ref{stationary_horizon} and \ref{stationary_naked}. Here, Figure 3 corresponds to a stationary black hole possessing a regular event horizon, whereas Fig. 4 illustrates a parameter choice for which no positive real root exists, leading to a horizonless configuration. These examples demonstrate that the stationary family likewise possesses a non-empty parameter space containing both horizon and horizonless geometries.

Upon rewriting the metric (\ref{332_1}) in the standard hyperscaling violating gauge $u=e^r$, the stationary spacetime takes the form of a rotating hyperscaling–violating geometry
\begin{equation}
ds^2=-u^{-2\theta/n} \left( -u^{2z} f(u) dt^2+\frac{du^2}{u^2 f(u)}+u^{2\xi} \left(d\phi+\frac{\omega}{r^2}dt\right)^2+\frac{u^2 dx_i dx^i}{\left( 1+\kappa \frac{\rho^2}{4}\right)^2}\right) \, ,
\end{equation}
\noindent with exponents
\begin{equation}
z=\beta_0 +\beta_3 \quad , \quad \theta = n(1-\beta_3)\quad, \quad \xi=\beta_1+\beta_3 \quad \mbox{and} \quad \beta_2=\beta_3 -1\, .
\end{equation}
\noindent Here $z$ denotes the dynamical exponent, $\theta$ the hyperscaling-violation exponent, and $\xi$ characterizes the anisotropic scaling of the rotational sector. Related Lifshitz-type black hole geometries in $R^{2}$ gravity were previously obtained in higher dimensions, indicating that quadratic curvature corrections naturally support anisotropic scaling solutions in appropriate parameter regimes. The rotating geometries obtained here generalize the nonrotating Lifshitz backgrounds discussed in \cite{sentorun} by incorporating homogeneous rotational deformations while preserving the characteristic anisotropic scaling
structure.

\section{Thermodynamic analysis}

We now turn to the thermodynamic properties of the Bianchi-type solutions. Although the horizon radius $r_+$, defined as the largest positive root of $f^2(r_+)=0$, cannot in general be obtained in closed form, the associated thermodynamic quantities can still be consistently analyzed. The black hole entropy is evaluated using Wald’s prescription \cite{wald},

\begin{equation} \label{321_20}
S_{\rm W}=-2 \pi \int_{S} \frac{\delta \mathcal L}{\delta R_{abcd}} n_{ab} n_{cd} , 
\end{equation}

\noindent where $\eta_{ab}$ denotes the binormal to the horizon, normalized as $\eta_{ab} \eta^{ab}=-2$. For the quadratic gravity model considered here, $f(R)=\frac{1}{2}\left( \alpha R^2 + R\right)+ \Lambda$ and the Wald entropy of a constant-curvature black hole reduces to
\begin{equation}
S_{\rm W}=\frac{A_H}{4}f'(R),
\end{equation}
where \(A_H\) denotes the horizon area. Since the solutions constructed in Section 4 satisfy the critical condition
\begin{equation} \label{fprimezero}
f^\prime(R)=1+2\alpha R=0,
\end{equation}
the entropy vanishes identically,
\begin{equation}
S_{\rm W}=0.
\end{equation}
For constant curvature black hole configurations arising in the presence of an $R^2$ correction, the entropy vanishes identically as a consequence of (\ref{fprimezero}). The vanishing of the Wald entropy results directly from condition (\ref{fprimezero}), which removes the effective Einstein--Hilbert contribution to the entropy functional. Similar behavior has been observed in several higher-derivative and critical gravity models, where black hole configurations may possess a finite surface gravity and temperature while the associated entropy vanishes.

The vanishing entropy can also be understood from the Noether-charge formulation of black hole thermodynamics developed by Wald. For diffeomorphism invariant theories, the conserved charge associated
with a Killing vector field $\xi$ is obtained from the Noether current constructed from the gravitational Lagrangian. In the present $R^{2}$ theory, the corresponding charge is proportional to the factor $f^\prime (R)$. Since the Bianchi-type solutions considered here satisfy the critical condition (\ref{fprimezero}), the Noether charge evaluated on the horizon vanishes identically. Consequently, the Wald entropy and the conserved
gravitational charge associated with the horizon both vanish despite the existence of a regular Killing horizon. A detailed derivation of the Noether-charge construction for the same $R^{2}$ gravity model was presented in \cite{sentorun}, and therefore will not be repeated here.

\noindent The surface gravity $K$ is defined on the Killing horizon generated by the vector field $\xi$. In terms of exterior calculus, it can be written as

\begin{equation} \label{321_22}
K=\frac{1}{2}\left(\ast(d \xi \wedge \ast d \xi)\right)^{(1/2)} .
\end{equation}

\noindent For a static, spherically symmetric spacetime with metric

\begin{equation} \label{321_23}
ds^{2}=-U(r)dt^{2}+\frac{dr^{2}}{V(r)}+R(r)d\Omega^{2} , 
\end{equation}

\noindent the normalized timelike Killing vector is $\xi=\partial_t$, and the surface gravity follows from the near-horizon limit

\begin{equation} \label{321_24}
K = \lim_{r\rightarrow r_{+}} \left(\frac{1}{2} \sqrt{\frac{V(r)}{U(r)}} U^{\prime}(r)\right).
\end{equation}
The Hawking temperature is then given by
\begin{equation} \label{321_25}
T=\frac{K}{2 \pi}.
\end{equation}
The surface gravity and the associated thermodynamic temperature of the Bianchi-type static black hole are given by
\begin{equation} \label{331_12}
K=\frac{\sqrt{n(n-1)}}{2} \lim_{r\rightarrow r_{+}} e^{(\beta_0-\beta_1)r}\left(f^2(r)\right)^{\prime},
\end{equation}
where $r=r_{+}$ denotes the location of the event horizon. The corresponding Hawking temperature is then obtained as
\begin{align}
T_{\mathrm{st}}
=
\frac{\sqrt{n(n-1)}}{4\pi}
\Big[
& p_{+}c_{1}
e^{(p_{+}+\beta _0-\beta _1)r_{+}}
+p_{-}c_{2}
e^{(p_{-}+\beta _0-\beta _1)r_{+}}
\nonumber\\
&+2\beta _1c_{3}
e^{(\beta _0+\beta _1)r_{+}}
+2(\beta _1-\beta _2)c_{4}
e^{(\beta _0+\beta _1-2\beta _2)r_{+}}
\Big].
\label{T_static}
\end{align}
The surface gravity and the associated thermodynamic temperature of the Bianchi-type stationary black hole in the $R^{2}$-corrected gravity theory, evaluated for vanishing entropy, are given by
\begin{equation} \label{332_13}
K=\frac{\sqrt{n(n-1)(n-2)}}{2}\lim_{r\rightarrow r_{+}} e^{(\beta_0-\beta_2)r}\left(f^2(r)\right)^{\prime}
\end{equation}
and
\begin{align}
T_{\mathrm{stat}}
=
\frac{\sqrt{n(n-1)(n-2)}}{4\pi}
\Bigg[
& p_{+}c_{1}
e^{(p_{+}+\beta _0-\beta _2)r_{+}}
+p_{-}c_{2}
e^{(p_{-}+\beta _0-\beta _2)r_{+}}
\nonumber\\
&+2\beta _2c_{3}
e^{(\beta _0+\beta _2)r_{+}}
+2(\beta _2-\beta _3)c_{4}
e^{(\beta _0+\beta _2-2\beta _3)r_{+}}
\nonumber\\
&+\frac{c_{5}}{r_{+}^{4}}
e^{(2\beta _1-\beta _0-\beta _2)r_{+}}
\left(
2(\beta _1-\beta _0)-\frac{4}{r_{+}}
\right)
\Bigg].
\label{T_stationary}
\end{align}

\noindent Static and stationary Bianchi black holes are characterized by a finite surface gravity and a nonvanishing temperature, while their entropy identically vanishes as a consequence of Eqs. (\ref{eq:Rcritical}) and (\ref{eq:alphacritical}) in $R^2$-corrected gravity. This behavior resembles that found for BTZ black holes in new massive gravity at the critical coupling, where the entropy vanishes while the Hawking temperature remains finite \cite{liu}. A direct consequence of the critical condition is that the Wald entropy
vanishes identically throughout the solution family. Since the specific heat is formally defined by
\[
C=T\left(\frac{\partial S}{\partial T}\right),
\]
the vanishing entropy implies
\[
C=0.
\]
Therefore, these solutions exhibit a degenerate thermodynamic response, in the sense that variations of the Hawking temperature are not accompanied by entropy variations within the critical sector. It should be emphasized, however, that the condition \(C=0\) alone does not establish thermodynamic stability. A complete stability analysis would require an independent investigation
of the canonical ensemble and the relevant free-energy landscape, which lies beyond the scope of the present work.

\noindent The Hawking temperatures corresponding to representative black hole solutions are summarized in Table 1. These values were obtained from the positive real roots of the metric function shown in Fig. 1 and Fig 3.

\begin{table}[H]
\centering
\caption{Comparison of thermodynamic parameters of solutions.}
\label{tab:termodinamik}
\begin{tabular}{|c|c|c|c|}
\textbf{Case} & \textbf{Parameters} & \textbf{Event Horizon ($r_+$)} & \textbf{Temperature ($T_H$)}  \\ \hline \hline
 & $c_1=6.304895$, $c_2=-6.772287$, & \\
 static& $n=5$, $\kappa=1$, $\Lambda=3.122499$, & 0.00177418 & 11.448 \\ 
 & $\beta_0=-0.303521$, $\beta_1=-2.391589$, & &  \\ 
 & $\beta_2=-1.626807$ & & \\ \hline
 &$c_1=-7.205257$, $c_2=-5.365793$, &  &   \\
stationary & $n=5$, $\kappa=1$, $\Lambda=-4.858181$, & 6.79995 & 2860.78 \\
 & $\beta_0=0.99813$, $\beta_1=0.31091$, & & \\ 
 & $\beta_2=0.056889$, $\beta_3=-0.20261$ & & \\ \hline
\end{tabular}
\end{table}

\section{Conclusion}

In this paper, we have investigated higher-dimensional Bianchi-type black hole solutions in $R^{2}$ gravity theory formulated within the framework of exterior differential forms. Working in the constant-curvature sector characterized by $R=-4\Lambda$ and $\alpha=1/(8\Lambda)$, we showed that the field equations admit a broad class of homogeneous but anisotropic geometries in arbitrary spacetime dimensions.

Two distinct families of solutions were constructed. The first consists of static Bianchi-type black holes supported by anisotropic scaling between the temporal, radial, and transverse directions. The second family corresponds to stationary configurations obtained through rotational deformations of the static backgrounds. In both cases, the resulting geometries are characterized by horizon structures associated with Bianchi-type homogeneous spaces and admit hyperbolic, planar, or spherical transverse sections according to the value of the curvature parameter $\kappa$.

A notable feature of the solutions is that, after an appropriate radial redefinition, both the static and stationary metrics can be expressed in a hyperscaling-violating form. The corresponding dynamical and hyperscaling-violation exponents are determined by the anisotropy parameters appearing in the metric, establishing a direct connection between Bianchi-type horizon geometries and anisotropic scaling backgrounds in higher-curvature gravity.

The thermodynamic analysis reveals an unconventional behavior. Owing to the critical condition $1+2\alpha R=0$, the Wald entropy and the associated Noether charge vanish identically, whereas the surface gravity and Hawking temperature remain finite. Consequently, these solutions provide explicit examples of constant-curvature black holes possessing finite Hawking temperature despite vanishing Wald entropy within the critical sector of \(R^2\)-corrected gravity. This behavior resembles that encountered in certain critical gravity models and highlights the nontrivial impact of quadratic curvature corrections on black-hole thermodynamics.

The solutions obtained here extend previously known Lifshitz-type backgrounds by incorporating homogeneous anisotropic horizon geometries and, in the stationary case, rotational deformations. More generally, the results demonstrate that $R^{2}$ gravity admits a considerably richer solution space than its Einstein counterpart, supporting exact black-hole configurations that combine anisotropy, hyperscaling violation, and nonstandard thermodynamic properties.

For future investigation, it would be worthwhile to examine whether similar Bianchi-type configurations persist in more general higher-curvature theories involving additional curvature invariants.

\section{Acknowledgements}

I would like to express my gratitude to Prof. Dr. Hakan Cebeci for his invaluable guidance and comments that enhanced both the clarity and depth of this study. This work is supported by Scientific Research Projects Commission of Eski\c{s}ehir Technical University under Grant no: 19ADP029.

\end{document}